\documentclass[aps,twocolumn,prl]{revtex4}
\usepackage{amsmath,amssymb,bm}
\usepackage{graphicx}
\usepackage{epstopdf}
\usepackage{latexsym}
\usepackage{times}
\usepackage{color}

\newcommand{\beq}{\begin{equation}}
\newcommand{\eeq}{\end{equation}}
\newcommand{\beqn}{\begin{eqnarray}}
\newcommand{\eeqn}{\end{eqnarray}}

\begin{document}
\title{Field theory for magnetic and lattice structure properties of $\mathrm{Fe_{1+y}Te_{1-x}Se_x}$}
\author{Cenke Xu}
\affiliation{Department of Physics, Harvard University, Cambridge
MA 02138, USA}
\author{Jiangping Hu}
\affiliation{Department of Physics, Purdue University, West
Lafayette, Indiana 47907, USA}
\date{\today}
\begin{abstract}

We study the magnetism and lattice distortion of
$\mathrm{Fe_{1+y}Te_{1-x}Se_x}$ with a general field theory
formalism, motivated by recent neutron scattering experiments.
Besides the Ising nematic order parameter which can be naturally
defined in the 1111 and 122 materials, We show that the collinear
order observed in $\mathrm{Fe_{1+y}Te}$ materials is stabilized by
an extra Ising order parameter which doubles the generalized unit
cell of the system. By tuning this Ising order parameter, one can
drive a transition between collinear and noncollinear spin density
wave orders, both of which are observed in the FeTe family. The
nature of the quantum and classical phase transitions are also
studied in this field theory formalism, with lattice strain tensor
fluctuation considered.

\end{abstract}

\maketitle

The Iron-superconductor, for its potential to shed new light on
the non-BCS type of superconductors, has attracted enormous
interests since early last year. So far most studies have been
focused on the 1111 materials and 122 materials. The 1111
materials have chemical formula $\mathrm{LnFeAsO}$, and
$\mathrm{Ln}$ represents one of the lanthanides; The 122 materials
have chemical formula $\mathrm{MFe_2As_2}$, and $\mathrm{M}$
usually represents one of the alkaline earths. At low temperature,
it is confirmed that both 1111 and 122 materials develop $(\pi,
0)$ spin density wave (SDW)
\cite{La1,LaFeAs2ndorder,Ba3,Sr1,Sr2,Sr3,Ca2}. Ever since its
discovery, the SDW in 1111 and 122 materials raised a debate about
whether this phenomenon should be described by an itinerant
electron picture with nesting fermi surface, or a local moment
Heisenberg model. This debate is partly due to the fact that the
ordering wave vector $(\pi, 0)$ connects two pieces of nearly
parallel fermi surfaces, and relatively small magnetic order
moment of 1111 materials. However, in the relatively less
extensively studied 11 material $\mathrm{FeTe}$ family, the answer
seems to be more clear, as the order wave vector of these
materials is exactly or close to $(\pi/2, \pi/2)$
\cite{FeTe,FeTeSe}, which is far away from the nesting wave vector
according to LDA calculation \cite{FeTeLDA}. Also, the magnetic
order moment can be as large as $2\mu_B$ \cite{FeTe}, which can be
naturally understood in terms of local moment picture.

In the low temperature phase of $\mathrm{Fe_{1+y}Te}$, besides the
spin density wave (SDW) with wave vector $(\pi/2, \pi/2)$, a
monoclinic lattice distortion \cite{FeTe} has also been
identified. By increasing the excess of Fe in this system, the
magnetic order becomes incommensurate \cite{FeTeSe}, with order
wave vector deviates from but still close to $(\pi/2, \pi/2)$.
Also, by replacing Te with Se, both SDW and lattice distortion are
suppressed. In Ref. \cite{andrei2008}, the authors used a local
moment Heisenberg model with the nearest, 2nd nearest and 3rd
nearest neighbor ($J_1 - J_2 - J_3$) interactions on the square
lattice to describe this system.  This simple model captures the
correct phases observed experimentally and is also supported by
results from LDA calculations\cite{Ma1}. But a model independent
general field theory formalism is also necessary, in order to for
instance explore the physics beyond the Heisenberg model, and to
study the nature of the phase transitions. In this work we will
systematically study the Ginzburg-Landau field theory for the
magnetism and lattice structures of FeTe compound and its relative
$\mathrm{Fe_{1+y}Te_{1-x}Se_x}$.

A general Ginzburg-Landau-Hertz-Millis theory for the 1111 and 122
materials was studied in Ref. \cite{qixu2008}. Motivated by
experimental facts, it was proposed in Ref.
\cite{kivelson2008,xms2008} that the true magnetic order in 1111
materials (say LaFeAsO) is developed through two separate steps:
first an anisotropic antiferromagnetic (AF) correlation is
developed at a relatively higher temperature, and then the long
range spin density wave (SDW) order is developed at lower
temperature. The anisotropic AF correlation described by an Ising
order parameter breaks the reflection symmetry of the lattice
along the axis $x = y$, and leads to a tetragonal-orthorhombic
lattice distortion. More specifically, this Ising order parameter
can be defined as $\Phi \sim \vec{\phi}_1\cdot\vec{\phi}_2$,
$\vec{\phi}_1$ and $\vec{\phi}_2$ are Neel order parameters on two
sublattices of the square lattice.

\begin{figure}
\centering
\includegraphics[width=2.4in]{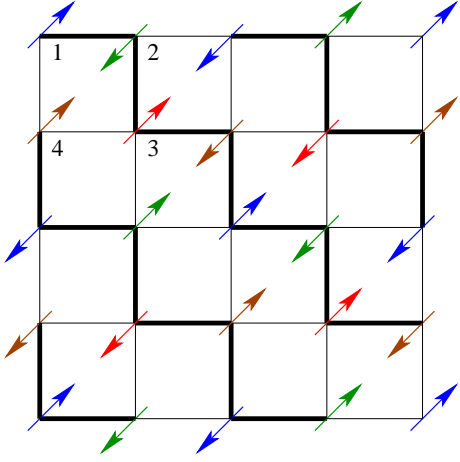}
\caption{The collinear magnetic order of FeTe family. The four
sublattices form Neel orders separately, and the thick lines
represent links with stronger AF correlations, which doubles the
unit cell defined by generalized translation operators
$T^\prime_x$ and $T^\prime_y$. It is possible that the anisotropic
AF correlations develop long range order before the actual SDW
pattern.} \label{patternFeTe}
\end{figure}

In the $\mathrm{FeTe}$ family, the SDW and lattice distortion are
more complicated, and presumably the anisotropic AF correlations
can also develop before the true long range SDW. The SDW order
pattern contains four Neel orders on four sublattices \cite{FeTe}
(Fig. \ref{patternFeTe}), $\vec{\phi}_a$ with $a = 1 - 4$. Two
Ising order parameters which obviously break the lattice
reflection symmetry are $\Phi_1 \sim \vec{\phi}_1 \cdot
\vec{\phi}_3$ and $ \Phi_2 \sim \vec{\phi}_2 \cdot \vec{\phi}_4$.
$\Phi_1$ and $\Phi_2$ can develop long range order before
$\vec{\phi}_a$ does, and the symmetry of $\Phi_i$ enables them to
distort the lattice from tetragonal to monoclinic once they
develop long range order. However, $\Phi_1$ and $\Phi_2$ are not
independent Ising order parameters $i.e.$ in the free energy there
should be a term which breaks the degeneracy between $\Phi_1\Phi_2
> 0$ and $\Phi_1 \Phi_2 < 0$. Later on we will
show in the ground state $\Phi_1 \Phi_2 < 0$, and the energy
barrier between states with $\Phi_1 \Phi_2 < 0$ and $\Phi_1 \Phi_2
> 0$ is estimated to be about $J_1^2 / J_3$ for the $J_1 - J_2 -
J_3$ model.

Let us first study the structure transitions, and tentatively
ignore the magnetic orders. In the high temperature symmetric
phase, the Fe-Te plane enjoys the following symmetries:
$\mathrm{P_z \otimes T_x}$, $\mathrm{P_z \otimes T_y}$,
$\mathrm{P_z \otimes P_x}$, $\mathrm{P_z \otimes P_y}$.
$\mathrm{P_z}$ and $\mathrm{P_x}$ are reflections $\hat{z}
\rightarrow -\hat{z}$ and $\hat{x} \rightarrow -\hat{x}$ with
origin located on one of the Fe atoms, $\mathrm{T_x}$ and
$\mathrm{T_y}$ are translations. Although the true unit cell of
the Fe-Te plane always involves two Fe atoms, one can define a one
Fe unit cell based on the generalized translations
$\mathrm{T_x}^\prime = \mathrm{P_z \otimes T_x}$ and
$\mathrm{T_y}^\prime = \mathrm{P_z \otimes T_y}$, and we will use
this generalized unit cell hereafter. The lattice distortion
observed in experiments\cite{FeTe} breaks all of the symmetries
including $\mathrm{T_x}^\prime$ and $\mathrm{T_y}^\prime$. The
lattice distortion favors the antiferromagnetic correlations along
the zig-zag stripes, and there are in total 4 different degenerate
zig-zag configurations. We can describe these zig-zag AF
correlation stripes with two Ising order parameters $\sigma_1$ and
$\sigma_2$. $\sigma_1 = \pm 1$ represents two degenerate zig-zag
correlation stripe patterns along the $x = y$ direction, and
$\sigma_2 = \pm 1$ represents two zig-zag correlation stripes
along the $x = - y$ direction. One can write down a
Ginzburg-Landau free energy for the coarse-grained modes of
$\sigma_a$ according to the symmetry of the system: \beqn
F_{\sigma} = \sum_{a = 1}^2 (\nabla_\mu \sigma_a)^2 + r \sigma_a^2
+ u_0 \sigma^4_a + u_{2} \sigma_1^2\sigma_2^2 + O(\sigma^6)
\label{freeenergysigma} \eeqn Here we need $u_2
> 2u_0$ to make sure that in the ordered phase
$\langle\sigma_1\rangle \langle\sigma_2\rangle = 0$, which is the
case in the real system.

Another more intuitive way to describe the $\sigma_a$ variables,
is to introduce complex field $\Psi = \sigma_1 + i\sigma_2$, and
the free energy Eq. \ref{freeenergysigma} can be rewritten as
\beqn F_{\Psi} = |\nabla_\mu \Psi|^2 + r |\Psi|^2 + g |\Psi|^4 +
g_4 (\Psi^4 + \Psi^{\ast 4}) + O(\Psi^6). \label{freeenergypsi}
\eeqn The free energy up to the forth order of $\sigma_a$ and
$\Psi$ enjoys an enlarged $Z_4$ symmetry, which can be viewed as
an $Z_4$ anisotropy of an XY or $\mathrm{U(1)}$ quantity $\Psi$.
It is well-known that the $Z_4$ anisotropy at the 3D XY transition
is irrelevant \cite{vicari2003}, so our first conclusion is that,
in the most simplified situation the finite temperature transition
of $\sigma_a$ is a 3D XY transition. In Eq. \ref{freeenergysigma}
and Eq. \ref{freeenergypsi} The terms at the 6th order or higher
will break this $Z_4$ symmetry, however, those terms will also be
irrelevant at the 3D XY transition, so we will ignore them from
now on.

\begin{figure}
\centering
\includegraphics[width=2.3in]{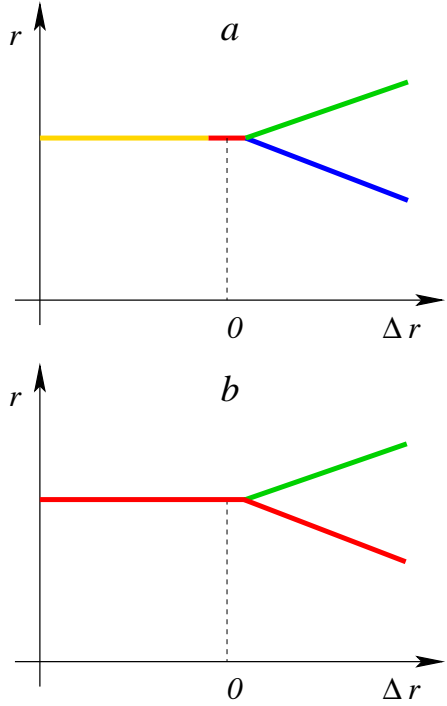}
\caption{The global phase diagrams for Eq. \ref{freeenergyfull}
and Eq. \ref{lagrangianfull} plotted against $r = r_1 + r_2$, and
$\Delta r = r_1 - r_2$. For Eq. \ref{freeenergyfull}, Fig. $a$
represents the case without considering lattice elasticity. In
this case the red line is a first order transition; the golden
line is a 3D XY Wilson-Fisher transition; the green line and blue
line are 3D Ising transitions for $\Phi$ and $\sigma_a$
respectively. For Eq. \ref{freeenergyfull}, Fig. $b$ represents
the phase diagram with coupling to the lattice strain tensor, the
3D XY transition and the 3D Ising transition of $\sigma_a$ become
first order, the 3D Ising transition of $\Phi$ becomes a mean
field transition. For Eq. \ref{lagrangianfull}, Fig. $a$ and Fig.
$b$ represent the case without and with considering itinerant
electron particle-hole excitations respectively. If a two
dimensional model is taken, the nature of the transitions is
almost the same as Eq. \ref{freeenergyfull}.} \label{phasediaFeTe}
\end{figure}

According to the symmetry, the Ising fields $\sigma_1$ and
$\sigma_2$ couple to the Ising field $\Phi \sim \Phi_1 \sim -
\Phi_2$ in the following manner: \beqn F_{\Phi, \sigma} = f_{\Phi,
\sigma} \Phi(\sigma_1^2 - \sigma_2^2). \label{fphisigma}\eeqn The
Ising field $\Phi$ describes the anisotropic AF correlation which
breaks no translational symmetry, but breaks the reflection
symmetries $\mathrm{P_z \otimes P_x}$ and $\mathrm{P_z \otimes
P_y}$. The full free energy reads: \beqn F &=& \sum_{a = 1}^2
(\nabla_\mu \sigma_a)^2 + r_1 \sigma_a^2 + (\nabla_\mu \Phi)^2 +
r_2 \Phi^2 \cr\cr &+& f_{\Phi,\sigma} \Phi(\sigma_1^2 -
\sigma_2^2) + \cdots. \label{freeenergyfull}\eeqn The ellipses
include all the quartic
terms. Both $r_1$ and $r_2$ are tuned by temperature. 
By minimizing Eq. \ref{freeenergyfull}, one can obtain the mean
field global phase diagram plotted against $r = r_1 + r_2$ and
$\Delta r = r_1 - r_2$ (Fig. \ref{phasediaFeTe}$a$). There are in
total three different regions: {\bf 1)}, when $r_1 \gg r_2$, there
are two transitions, with one Ising transition for $\Phi$ and
another Ising transition for one of the $\sigma_a$, since in this
region $\Phi$ has a much stronger tendency to order compared with
$\sigma_a$; {\bf 2)}, when $r_1 \sim r_2$, there is an
intermediate region with first order transition, and $\Phi$ and
$\sigma_a$ will order at the same temperature. The first order
nature of this transition is due to the cubic coupling term
$F_{\Phi,\sigma}$. {\bf 3)}, when $r_1 \ll r_2$ there is still one
single transition, but this transition is second order. This
transition can be studied by integrating out $\Phi$ from free
energy Eq. \ref{freeenergyfull}, and the resultant free energy for
$\sigma_a$ is the same as Eq. \ref{freeenergypsi}, which describes
an 3D XY transition as was discussed previously.

So far we have ignored the lattice elasticity, which will
potentially change the nature of the transitions discussed above
\cite{qixu2008}. In region {\bf 1)} discussed in the previous
paragraph, the Ising field $\Phi$ will couple to the uniaxial
shear strain field of the lattice: \beqn F_{\Phi, u} \sim \Phi
(\partial_x u_x -
\partial_y u_y), \eeqn $\vec{u}$ represents the local displacement
vector of the lattice. By integrating out the displacement vector
$\vec{u}$, we can show that the fluctuation of the displacement
vector $\vec{u}$ will increase the effective spatial dimension at
the transition of $\Phi$, and drive the transition of $\Phi$ mean
field like, as was studied carefully in Ref. \cite{qixu2008}. At
the mean field transition the specific heat curve will have a
discontinuity instead of a peak, as was observed in electron-doped
122 materials \cite{Co1}. Unlike the field $\Phi$, in region {\bf
1)} the field $\sigma_a$ that breaks the translational symmetry
will only couple to the bulk strain field in the following way:
\beqn F_{\sigma, u} = f_{\sigma, u} |\sigma_a|^2 (\nabla_x u_x +
\nabla_y u_y + \lambda \nabla_z u_z). \eeqn We can estimate the
scaling dimension of the coupling constant $f_{\sigma, u}$ as the
following: The scaling dimension of $|\sigma_a|^2$ at the 3D Ising
transition is $\Delta[|\sigma_a|^2] = 3 - 1/\nu$, and the scaling
dimension of displacement vector $\vec{u}$ is $\Delta[\vec{u}] =
1/2$. Therefore the dimension of $f_{\sigma, u}$ is
$\Delta[f_{\sigma, u}] = - 3/2 + 1/\nu = \alpha_{ising}/2 > 0$. So
this coupling will lead to a relevant perturbation at the 3D Ising
transition of $\sigma_a$. This relevant perturbation will likely
drive the transition first order \cite{dh}. This argument of
scaling dimension is essentially the same as the well-known Harris
criterion in the disordered system.


In region {\bf 3)}, as was shown before, after integrating out
$\Phi$, the transition is described free energy Eq.
\ref{freeenergypsi}, which describes a 3D XY transition. The 3D XY
transition is also affected by the lattice elasticity. First of
all, $|\Psi|^2$ is also coupled to the bulk strain field $
F_{\Psi, u} \sim |\Psi|^2 (\nabla_x u_x + \nabla_y u_y +
\lambda^\prime \nabla_z u_z)$, because $\alpha_{xy} = 2/\nu - D <
0$ this coupling will not introduce any relevant perturbations.
However, $\sigma^2_1 - \sigma^2_2 \sim \mathrm{Re}[\Psi^2]$ will
couple to the shear strain field $
\partial_x u_x - \partial_y u_y$. To check whether this coupling
is relevant or not, we just need to compare the scaling dimension
of $\mathrm{Re}[\Psi^2]$ to $D / 2$ at the three dimensional XY
transition. At the 3D XY critical point, the scaling dimension
$\Delta[\mathrm{Re}[\Psi^2]] = 1.234 < 3/2$ \cite{vicari2003},
hence this coupling will indeed generate relevant perturbation at
the 3D XY transition, which likely drives the transition a first
order one. Therefore after considering the lattice elasticity, the
global phase diagram is modified to Fig. \ref{phasediaFeTe}$b$.

Now let us consider the SDW order, described by four Neel order
parameters $\vec{\phi}_a$ with $a = 1 - 4$, and let us ignore the
Ising fields $\sigma_a$ first. Besides the ordinary kinetic terms
in the free energy, several linear spatial derivative terms are
also allowed by the symmetry of the system, therefore a most
general free energy for the Neel order parameter $\vec{\phi}_a$
reads: \beqn F_{\vec{\phi}} &=& \sum_{a = 1}^4 J_3(\nabla_\mu
\vec{\phi}_a)^2 + \beta \vec{\phi}_1 \cdot \nabla_x \vec{\phi}_2 +
\beta \vec{\phi}_4\cdot \nabla_x \vec{\phi}_3 \cr\cr &-& \beta
\vec{\phi}_2 \cdot \nabla_y \vec{\phi}_3 - \beta \vec{\phi}_1
\cdot \nabla_y \vec{\phi}_4 \cr\cr &-& g\Phi_1\vec{\phi_1} \cdot
\vec{\phi}_3 - g \Phi_2\vec{\phi}_2 \cdot \vec{\phi}_4. \eeqn The
$\hat{z}$ direction dispersion has been ignored. This theory can
also be viewed as a low energy field theory for the lattice $J_1 -
J_2 - J_3$ Heisenberg model, with $\beta \sim J_1$, and $g \sim
J_2^2/J_3$. If $\Phi_1 \Phi_2 < 0$, An infinitesimal $J_1$
inevitably drives the system to an incommensurate order, with
ordering wave vector $\vec{Q} = (\pi/2 + \delta q, \pi/2 + \delta
q)$, $\delta q \sim J_1/J_3$. However, if $\Phi_1 \Phi_2 > 0$, the
minima of the dispersion remains at $\vec{Q} = (\pi/2, \pi/2)$ for
small enough $J_1$. The energy difference between the two states
with $\Phi_1 \Phi_2 < 0$ and $\Phi_1 \Phi_2 > 0$ is about
$J_1^2/J_3$, or in other words there is an effective coupling
$\Phi_1 \Phi_2$ in the free energy, which is proportional to
$J_1^2 / J_3$.

In the phase with $\Phi_1\Phi_2 < 0$ The collinear commensurate
order is stabilized by coupling the SDW order parameter to the
Ising fields $\sigma_a$. Again, the symmetry of the system allows
the following coupling: \beqn F_{\sigma, \vec{\phi}} &=& \gamma
\sigma_1 (\vec{\phi}_1 \cdot \vec{\phi}_2 + \vec{\phi}_2 \cdot
\vec{\phi}_3 - \vec{\phi}_3 \cdot \vec{\phi}_4 - \vec{\phi}_1
\cdot \vec{\phi}_4) \cr\cr & + & \gamma \sigma_2 (\vec{\phi}_1
\cdot \vec{\phi}_2 - \vec{\phi}_2 \cdot \vec{\phi}_3 -
\vec{\phi}_3 \cdot \vec{\phi}_4 +\vec{\phi}_1 \cdot \vec{\phi}_4).
\label{freeenergyspin}\eeqn After diagonalizing the quadratic part
of the entire free energy (Eq. \ref{freeenergyspin}), we can see
that if $\Phi_1
> 0$ and $\Phi_2 < 0$ the system favors $\langle\sigma_2\rangle \neq
0$ while $\langle\sigma_1\rangle = 0$. Therefore there is indeed
an effective coupling $F_{\Phi, \sigma}$, as Eq. \ref{fphisigma}.

Whether the system develops incommensurate or commensurate order,
depends on the competition between the terms with linear spatial
derivatives, and the terms with coupling to $\sigma_a$. When
$\gamma \langle\sigma_a\rangle $ dominates $J_1^2/J_3$, the
commensurate collinear order is stabilized, otherwise the system
develops the incommensurate coplanar order. Assuming $\Phi =
\Phi_1 > 0$, and hence at low temperature $\sigma_2$ orders while
$\sigma_1$ remains disordered, a transition between incommensurate
and commensurate orders can be driven by tuning $\sigma = \langle
\sigma_2\rangle$. Experimentally this transition can be achieved
by changing the excess of Fe \cite{FeTeSe}. If we diagonalize the
quadratic part of the SDW free energy, the mode with the lowest
minimum takes the following form: \beqn E(\vec{q}) \sim -
g|\langle \Phi \rangle| + J_3 (q_x^2 + q_y^2) -
\sqrt{\gamma^2\sigma^2 + \beta^2 (q_x - q_y)^2}. \eeqn One can
check that by tuning $\sigma$, there is a transition between
commensurate and incommensurate order, and the incommensurate wave
vector is a continuous function of $\sigma$. The commensurate
collinear SDW order has ground state manifold $S^2$ times lattice
transformation degeneracy, while the incommensurate coplanar SDW
order has ground state manifold $S^3/Z_2$. The transition between
the these two phases can be viewed as condensation of an U(1)
degree of freedom \cite{xusachdev2008}, because the manifold $S^3$
is locally $S^2 \times S^1$. However, as was shown in Ref.
\cite{xusachdev2008}, the condensation of this U(1) degree of
freedom does not belong to the XY universality class, because the
spin-wave fluctuation of the collinear SDW order parameter at the
transition will again effectively increase the spatial dimension
of the XY transition, therefore the collinear/coplanar phase
transition always belongs to the mean field universality class,
for both 3D classical and 2+1d quantum cases \cite{xusachdev2008}.

It is also interesting to study the quantum phase transitions of
$\sigma_a$ and $\Phi$ at zero temperature, and we will take a
simple two dimensional model to emphasize the physics within the
Fe-Te plane, and also partly motivated by the fact that the band
structure calculated by LDA has a much weaker $\hat{z}$ direction
dispersion for $\mathrm{FeTe}$ and $\mathrm{FeSe}$ compared with
the 122 compounds \cite{FeTeLDA}. The Lagrangian describing the
quantum phase transitions is very similar to the free energy Eq.
\ref{freeenergyfull}: \beqn L &=& \sum_{a = 1}^2(\omega^2 +
v^2_{\sigma}q^2 + r_1) |\sigma_{a, \omega, q}|^2 \cr\cr &+&
(\omega^2 + v^2_{\Phi}q^2 + r_2)|\Phi_{\omega, q}|^2 \cr\cr &+&
f_{\Phi,\sigma} \Phi(\sigma_1^2 - \sigma_2^2) + \cdots
\label{lagrangianfull}\eeqn The phase diagram is actually quite
similar to Fig. \ref{phasediaFeTe}, with $r$ tuned by replacing Te
with Se in $\mathrm{Fe_{1+y}Te}$, as shown experimentally
\cite{FeTe,FeTeSe}. Phase diagram Fig. \ref{phasediaFeTe}$a$
corresponds to the case without fermi pockets, or phase
transitions in a superconductor with fully gapped fermi surface,
and the nature of the phase transitions described by Eq.
\ref{lagrangianfull} is the same as the phase transitions in Eq.
\ref{freeenergyfull}.

The fermi surface of the 11 materials have been observed by ARPES
\cite{FeTearpes}, and the structure qualitatively agrees with the
LDA calculation. If the fermi surfaces are not gapped, the
coupling between the bosonic order parameters $\sigma_a$ and
$\Phi$ with the particle-hole excitations around the fermi surface
will change the nature of the quantum phase transitions, or more
specifically change the dynamical exponent $z$. Phase diagram Fig
.\ref{phasediaFeTe}$b$ corresponds to the case with ungapped fermi
pockets. The green line becomes a $z = 3$ quantum phase
transition, as discussed in Ref. \cite{xms2008,kivelson2001},
because the order $\Phi$ carries zero momentum, and the
particle-hole excitation around the fermi surface will induce a
decay term of $\Phi$: \beqn L_{\Phi, d} \sim \frac{|\omega|}{q}
|\Phi_{\omega, q}|^2. \eeqn A similar decay term does not exist at
the quadratic level in the Lagrangian of $\sigma_a$, because
$\sigma_a$ carries momentum $(\pi/2, \pi/2)$. Based on the LDA
calculation \cite{FeTeLDA} and also ARPES measurements
\cite{FeTearpes}, the momentum $(\pi/2, \pi/2)$ does not connect
two pieces of the fermi surfaces. However, the particle-hole
excitation can also modify the quartic terms of the Lagrangian Eq.
\ref{lagrangianfull}. The coupling between $|\sigma_a|^2$ and
fermion density will lead to a term \beqn L_{\sigma, d} \sim
|\sigma_a|^2_{\omega,
\vec{q}}\frac{|\omega|}{q}|\sigma_a|^2_{-\omega, -\vec{q}}, \eeqn
which is a relevant perturbation at the 3D Ising transition due to
the positive $\alpha_{ising}$. On the left side of Fig.
\ref{phasediaFeTe}$b$ with $\Delta r < 0$, the transition was a 3D
XY transition without fermi pockets. The Fermi pockets will
generate a term $\mathrm{Re}[\Psi^2]_{\omega,
\vec{q}}\frac{|\omega|}{q}\mathrm{Re}[\Psi^2]_{-\omega,
-\vec{q}}$, which is again relevant at the 3D XY transition.
Therefore the red line in Fig .\ref{phasediaFeTe}$b$ is a first
order transition for a 2 dimensional quantum model.

In summary, we have shown that $\mathrm{Fe_{1+y}Te}$ materials are
characterized by two Ising order parameters: one Ising nematic
order which has been found in 1111 and 122 materials and the other
Ising order which doubles the unit cell of systems. We have also
calculated the phase diagram  and  the nature of phase
transitions. Unlike long-range SDW  orders which are easily
destroyed by replacing $\mathrm{Te}$ with $\mathrm{Se}$ in
$\mathrm{FeTe_{1-x}Se_x}$, the Ising orders may survive in larger
doping regions\cite{kivelson2008,xms2008} and coexist with
superconducting states.  Experimentally, it is possible to detect
the Ising orders from their broken symmetries, for example, using
polarized light in angle resolved photoemission spectra.



\paragraph*{Acknowledge} We thank B.A. Bernevig for useful
discussions. JPH is supported by the NSF under grant No.
PHY-0603759.


\vskip -0.2in
\begin{thebibliography}{99}

\vskip -0.2in

\bibitem{La1}  Clarina de la Cruz, Q. Huang, J. W. Lynn, Jiying Li, W. Ratcliff II,
J. L. Zarestky, H. A. Mook, G. F. Chen, J. L. Luo, N. L. Wang,
Pengcheng Dai, Nature {\bf 453}, 899 (2008).

\bibitem{LaFeAs2ndorder} M. A. McGuire, A. D. Christianson, A. S. Sefat, B. C. Sales, M. D. Lumsden,
R. Jin, E. A. Payzant, D. Mandrus, Y. Luan, V. Keppens, V.
Varadarajan, J. W. Brill, R. P. Hermann, M. T. Sougrati, F.
Grandjean, G. J. Long, Phys. Rev. {\bf 78}, 094517 (2008).

\bibitem{Ba3}  Q. Huang, Y. Qiu, Wei Bao, J.W. Lynn, M.A. Green, Y. Chen, T. Wu, G. Wu, X.H. Chen,
arXiv:0806.2776 (2008).

\bibitem{Sr1}  C. Krellner, N. Caroca-Canales, A. Jesche, H. Rosner, A. Ormeci, C. Geibel,
Phys. Rev. B {\bf 78}, 100504(R) (2008).

\bibitem{Sr2}  J.-Q. Yan, A. Kreyssig, S. Nandi, N. Ni, S. L. Bud'ko, A. Kracher,
R. J. McQueeney, R. W. McCallum, T. A. Lograsso, A. I. Goldman, P.
C. Canfield, Phys. Rev. B. {\bf 78}, 024516 (2008).

\bibitem{Sr3}  Jun Zhao, W. Ratcliff II, J. W. Lynn, G. F. Chen,
J. L. Luo, N. L. Wang, Jiangping Hu, Pengcheng Dai, Phys. Rev. B
{\bf 78}, 140504(R) (2008).

\bibitem{Ca2} A.I. Goldman, D.N. Argyriou, B. Ouladdiaf, T. Chatterji, A. Kreyssig,
S. Nandi, N. Ni, S. L. Bud'ko, P.C. Canfield, R. J. McQueeney,
Phys. Rev. B {\bf 78}, 100506(R) (2008).

\bibitem{Co1}  Jiun-Haw Chu, James G. Analytis, Chris Kucharczyk, Ian R. Fisher,
arXiv:0811.2463 (2008).

\bibitem{andrei2008} Chen Fang, B. Andrei Bernevig, Jiangping Hu, arXiv:0811.1294 (2008).

\bibitem{Ma1} Fengjie Ma, Wei Ji, Jiangping Hu, Zhong-Yi Lu and Tao
Xiang, Arxiv:0809.4732 (2008).

\bibitem{FeTe} Shiliang Li, Clarina de la Cruz, Q. Huang, Y. Chen, J. W. Lynn, Jiangping Hu,
Yi-Lin Huang, Fong-chi Hsu, Kuo-Wei Yeh, Maw-Kuen Wu, Pengcheng
Dai, Phys. Rev. B 79, 054503 (2009).

\bibitem{FeTeSe} Wei Bao, Y. Qiu, Q. Huang, M.A. Green, P. Zajdel, M.R. Fitzsimmons, M. Zhernenkov,
M. Fang, B. Qian, E.K. Vehstedt, J. Yang, H.M. Pham, L. Spinu,
Z.Q. Mao,  arXiv:0809.2058 (2008).

\bibitem{vicari2003} Pasquale Calabrese, Andrea Pelissetto, Ettore
Vicari, cond-mat/0306273, (2003).

\bibitem{kivelson2008} Chen Fang, Hong Yao, Wei-Feng Tsai, JiangPing Hu, Steven A. Kivelson,
Phys. Rev. B 77, 224509 (2008).

\bibitem{coleman1990} P. Chandra, P. Coleman, and A. I. Larkin,
Phys. Rev. Lett 64, 88 (1990).

\bibitem{xms2008} Cenke Xu, Markus Mueller, Subir Sachdev,
Phys. Rev. B 78, 020501(R) (2008).

\bibitem{qixu2008} Yang Qi and Cenke Xu,
arXiv:0812.0016 (2008).

\bibitem{xusachdev2008} Cenke Xu and Subir Sachdev,
Phys. Rev. B 79, 064405 (2009).

\bibitem{vicari2003} Pasquale Calabrese, Andrea Pelissetto, Ettore
Vicari, cond-mat/0306273, (2003).

\bibitem{kivelson2001} Vadim Oganesyan, Steven Kivelson, Eduardo Fradkin,
Phys. Rev. B 64, 195109 (2001).

\bibitem{dh} D.~Bergman and B.~I.~Halperin, Phys. Rev. B. {\bf 13}, 2145 (1976).

\bibitem{FeTeLDA} Alaska Subedi, Lijun Zhang, David J. Singh, Mao-Hua Du,
Phys. Rev. B 78, 134514 (2008).

\bibitem{FeTearpes} Y. Xia, D. Qian, L. Wray, D. Hsieh, G.F. Chen, J.L. Luo, N.L. Wang, M.Z. Hasan,
arXiv:0901.1299 (2009).

\end{thebibliography}
\end{document}